# Electron effective mobility in strained Si/Si$_{1-x}$Ge$_x$ MOS devices using Monte Carlo simulation

V. Aubry-Fortuna, P. Dollfus, S. Galdin-Retailleau

Institut d'Electronique Fondamentale, CNRS UMR 8622, Bât. 220, Université Paris-Sud, 91405 Orsay cedex, France.

E-mail : valerie.fortuna@ief.u-psud.fr

## Abstract

Based on Monte Carlo simulation, we report the study of the inversion layer mobility in n-channel strained Si/ Si$_{1-x}$Ge$_x$ MOS structures. The influence of the strain in the Si layer and of the doping level is studied. Universal mobility curves $\mu_{\text{eff}}$ as a function of the effective vertical field $E_{\text{eff}}$ are obtained for various state of strain, as well as a fall-off of the mobility in weak inversion regime, which reproduces correctly the experimental trends. We also observe a mobility enhancement up to 120 % for strained Si/ Si$_{0.70}$Ge$_{0.30}$, in accordance with best experimental data. The effect of the strained Si channel thickness is also investigated: when decreasing the thickness, a mobility degradation is observed under low effective field only. The role of the different scattering mechanisms involved in the strained Si/ Si$_{1-x}$Ge$_x$ MOS structures is explained. In addition, comparison with experimental results is discussed in terms of SiO$_2$/ Si interface roughness, as well as surface roughness of the SiGe substrate on which strained Si is grown.

## Keywords

## PACS codes

# 1. Introduction

The use of strained-Si channel pseudomorphically grown on a SiGe virtual substrate is becoming a promising way to accelerate the improvement of CMOS performance. In strained-Si surface channel configuration with gate oxide, both n-MOS and p-MOS transistors should strongly benefit from strain-induced enhancement of carrier transport. The biaxial tensile strain introduces splitting of degenerate bands [1] which results, for both electrons and holes, in smaller in-plane conduction mass and reduced intervalley scattering thereby yielding improved carrier velocity. In the case of electrons this effect has been clearly shown from mobility measurement and calculation in SiGe-Si-SiGe quantum wells [2-4] and it has been used for designing high-performance MODFET with low noise figure and high cut-off and maximum oscillation frequencies [5-6]. Now efforts are made to transfer this advantage in CMOS technology on either bulk or insulating substrate [7-10]. It has been demonstrated that the effective electron mobility in MOS structures can be substantially enhanced using tensile strained-Si channel grown on SiGe buffer layer [7, 11, 12, 13]. Compared to unstrained Si structures, a 120% improvement has been recently reported [13]. The aim of this article is to carefully analyse the effect of strain on electron transport in the Si inversion layer for MOS structures designed on bulk substrate. This theoretical study is based on the particle Monte Carlo method for solving the Boltzmann transport equation.

Many works have been dedicated to the study of transport in bulk Si MOSFET inversion layer, both experimentally [14-16] and theoretically [17-20]. Since pioneering work of Sabnis and Clemens this type of transport is commonly characterized by the curves of carrier mobility plotted as a function of effective vertical field defined by [14]

$$E_{eff} = \frac{Q_b + \eta\, Q_i}{\varepsilon_{Si}} \quad (1)$$

where $Q_b$ and $Q_i$ are the bulk depletion and inversion charges per unit area, respectively, $\varepsilon_{Si}$ is the Si dielectric constant and $\eta$ is an empirical parameter. For state-of-the-art bulk-Si MOSFET it has been clearly established that by taking $\eta$ equal to 0.5 for electrons the mobility-field curves obey a "universal mobility" law independently of substrate doping [15, 16]. More recently, a similar universal behaviour has been observed for strained-Si MOSFET [11]. Similarly, our present work is based on the calculation of mobility versus electric field in MOS structures.

Our device Monte Carlo code includes a 3D Poisson solver self-consistently coupled to the Monte Carlo algorithm [21]. For the present work limited to effective mobility extraction, the Poisson coupling is disabled. According to the gate bias, the 1D Poisson's equation is solved in the MOS capacitor prior to the Monte Carlo computation which is then made under frozen vertical field and uniform lateral driving field. This non-self-consistent approach may include quantization effects through coupling to 1D Schrödinger's equation [4]. **Such** **quantization effects are expected to be strong in future bulk devices because the necessary use of high doping to reduce short channel effects may induce high transverse electric field. In this regard device simulation should include these effects. However, device simulators able to accurately describe scattering mechanisms (MC simulators) together with quantization effects through self-consistent coupling of the Poisson and Schrödinger equations have not yet reach a high level of maturity. Thus, despite a loss of accuracy, it may be considered that a 3D description of the carrier gas in the channel is still useful and relevant provided it is able to correctly reproduce experimental results. That is why quantization is not taken into account in this work. The empirical roughness scattering model presented below has been developed for a 3D gas in such a way that it can be used for self-consistent device simulation which does not include quantization yet.**

To correctly model the transport in inversion layer, a careful description of both roughness scattering and impurity scattering is required. The former dominates the transport at high effective field $E_{eff}$ and the latter strongly varies as a function of $E_{eff}$ as a result of the carrier density dependence of the screening effect. The impurity scattering model has been modified to include the local carrier density in the evaluation of the screening function. Surface roughness scattering is treated with the widely used technique for 3D electron gas which consists of an empirical combination of diffusive and specular reflections at oxide interface [22-24]. However, to recover the universal mobility on a large range of substrate doping and effective field, we show that the fraction of diffusive scattering should not be a constant but a function of $E_{eff}$.

The theoretical background of the Monte Carlo model is presented in Sec. 2. The results of electron transport in bulk-Si MOS structures are described in Sec. 3 and compared with experimental results. The transport in strained Si structures is then carefully analyzed in Sec. 4 as a function of strain and Si thickness.

## 2. Monte Carlo Model for electrons in inversion layer

Here in Sect. 2 we give details of the Monte Carlo model used to describe the electron transport in Si, SiGe and strained Si in the case of MOSFET inversion layer.

### 2.1. Conduction-band structure of strained Si/ Si$_{1-x}$Ge$_x$

In this work, the conduction band of unstrained Si consists of six non parabolic $\Delta$ valleys located along the [100] directions at 85% of the Brillouin zone edge. The equi-energy surfaces are ellipsoidal with the transverse effective mass $m_t = 0.1905\ m_0$ and the longitudinal effective mass $m_l = 0.9163\ m_0$ ($m_0$ being the free electron mass). The non parabolicity coefficient $\alpha$ is assumed to be 0.5. For Si(100) inversion layer, it is convenient to denote normal (or $\Delta_2$) valleys the two valleys that have the longitudinal axis normal to the plane of growth and to denote parallel (or $\Delta_4$) valleys the four valleys that have the longitudinal axis in this plane.

For strained Si inversion layers on a (100)-$Si_{1-x}Ge_x$ pseudosubstrate, the normal valleys shift down with respect to the parallel ones, therefore favoring electron transport with a transverse mass in the plane of growth. In bulk $Si_{1-x}Ge_x$ (for $x \leq 0.3$), the conduction-band structure remains Si-like with six $\Delta$ valleys assumed to be undistorted in the presence of Ge. We assume here that, at first order, transverse and longitudinal masses in tensile-strained Si and bulk $Si_{1-x}Ge_x$ remain unchanged by strain or in the presence of Ge [25, 26]. In addition, the effect of strain in Si is included in the splitting energy $\Delta E_S$ between the two lowered normal valleys and the four raised parallel valleys. $\Delta E_S$ is given as a function of the Ge content $x$ in $Si_{1-x}Ge_x$ by $\Delta E_S = 0.68\,x$ (eV) [27, 28]. The presence of strain also reduces the Si band-gap, which leads to a conduction band-offset $\Delta E_c$ at the Si/ $Si_{1-x}Ge_x$ interface given by $\Delta E_c = 0.55\,x + 0.1\,x^2$ (eV) [27, 28].

## 2.2. Scattering mechanisms

The scattering mechanisms included in the Monte Carlo model of electron transport in Si are acoustic intravalley scattering, intervalley scattering via three $f$ and three $g$ phonons, ionized impurity scattering, surface-roughness scattering at $SiO_2$/Si interface and, for SiGe only, alloy scattering.

### 2.2.1. Phonon scattering

The acoustic intravalley phonon scattering is treated as an elastic process. By neglecting anisotropic effects the scattering rate in $\Delta$ valleys is given by [29]

$$\lambda_{ac}(E) = \frac{\sqrt{2}}{\pi} \frac{k_B T\, m_D^{3/2}\, D_{ac}^2}{\hbar\, \rho\, u^2} \sqrt{E(1+\alpha E)}\,(1+2\alpha E) \tag{2}$$

where $E$ is the electron kinetic energy, $m_D = \left(m_t^2\, m_l\right)^{1/3}$ is the density of states effective mass, $u = (u_l + 2u_t)/3$ is the average sound velocity, $k_B$ the Boltzmann constant, $T$ the lattice temperature, $\rho$ is the material density and $D_{ac}$ is the acoustic deformation potential.

The $\Delta$-$\Delta$ intervalley transitions are generally treated via the usual zeroth-order transition matrix, which yields to the following expression of scattering rate [29]

$$\begin{aligned}\lambda_{iv_0}(E) = &\frac{Z_{iv}}{\sqrt{2}\,\pi} \frac{m_D^{3/2}\, D_0^2}{\hbar^2 \rho\, \hbar\omega_{iv}} \left[N_p + \frac{1}{2} - \frac{\sigma}{2}\right] \sqrt{E + \sigma\hbar\omega_{iv} + \Delta E_{iv}} \times \\ &\times \sqrt{1+\alpha(E+\sigma\hbar\omega_{iv}+\Delta E_{iv})}\,\left[1+2\alpha(E+\sigma\hbar\omega_{iv}+\Delta E_{iv})\right]\end{aligned} \tag{3}$$

where $\hbar\omega_{iv}$ is the phonon energy, $D_0$ is the corresponding deformation potential, $N_p$ is the phonon number, $\Delta E_{iv}$ is the intervalley energy transition, $Z_{iv}$ is the number of possible final valleys, $\sigma$ is equal to $-1$ in the case of phonon absorption and to $+1$ in the case of phonon emission.

However, according to selection rules, the transitions with low energy phonons, *i.e.* $f_1$ (TA), $g_1$ (TA), and $g_2$ (LA) are forbidden via zero order process. Ferry proposed they should be considered by expanding the transition matrix to the first-order in the phonon wave vector without transgressing the selection rules [30]. The deformation potential $D_1$ is then expressed in eV instead of eV/cm for the coupling constant $D_0$ of usual zero order process. The wave vector dependence of the scattering rate becomes a simple energy dependence using the isotropic approximation $E(E+\alpha E)=\hbar^2 k^2/2m_D$. If $E'$ stands for the final energy $E'=E+\sigma\hbar\omega_{iv}+\Delta E_{iv}$ the complete expression of the first-order scattering rate is finally

$$\lambda_{iv_1}(E) = \frac{\sqrt{2}\, Z_{iv}}{\pi} \frac{m_D^{5/2}\, D_1^2}{\hbar^4 \rho\, \hbar\omega_{iv}} \left[N_p + \frac{1}{2} - \frac{i}{2}\right] \times$$
$$\times (1+2\alpha E')\sqrt{E'(1+\alpha E')}\left[E'(1+\alpha E') + E(1+\alpha E)\right] \quad (4)$$

We have adopted this approach to treat Δ-Δ intervalley scattering through $f_1$, $g_1$ and $g_2$ phonons while scattering through $f_2$, $f_3$ and $g_3$ phonons are considered via the usual zeroth-order process [4, 31]. A similar approach is used by other authors [32-34]. The material and phonon parameters used in the calculation are listed in Table I.

### 2.2.2. Impurity scattering

Within the Born approximation, impurity scattering is treated via the screened Coulomb potential using the momentum-dependent screening length $L$ defined by [35]

$$\frac{1}{L^2} = \frac{e^2\, n(\mathbf{r})}{\varepsilon_S\, k_B T} F(\xi) = q_D^2\, F(\xi) \quad (5)$$

where $q_D$ is the inverse of the Debye-Hückel screening length, $n(\mathbf{r})$ is the local carrier density, $\varepsilon_S$ is the dielectric permittivity and the normalised variable $\xi$ is defined by

$$\xi^2 = \frac{\hbar^2}{8 e\, m_D\, k_B T}\, q^2 \quad (6)$$

where $\mathbf{q} = \mathbf{k} - \mathbf{k}'$ is the wave vector involved in scattering event from an initial state $\mathbf{k}$ to a final state $\mathbf{k}'$. The screening function $F(\xi)$ has been derived for nondegenerate semiconductors [35, 36] and may be conveniently rewritten as [37]

$$F(\xi) = \frac{1}{\xi}\exp(-\xi^2)\int_0^\xi \exp(x^2)dx \quad (7)$$

which is a tabulated function [38]. Considering the scattering angle $\theta$, the impurity scattering rate for an electron in state $\mathbf{k}$ of energy $E$ is given by

$$\lambda_{imp}(k) = \frac{Z^2\, e^4}{\sqrt{2}\,\pi\,\hbar^4\,\varepsilon_S^2} m_D^{3/2}\, N_{imp}(1+2\alpha E)\sqrt{E(1+\alpha E)}\int_{-1}^{1}\frac{d(\cos\theta)}{\left(q^2 + q_D^2\, F(\xi)\right)^2} \quad (8)$$

where $N_{\text{imp}}$ is the impurity concentration and $Z$ is the number of charge units of the impurity. In the integral, the terms $q$ and $\xi$ depend on the angle $\theta$ through $q^2 = 2k^2(1-\cos\theta)$. Again, the isotropic approximation of the dispersion relation may be conveniently used to make $\lambda_{\text{imp}}(k)$ an energy-dependent function $\lambda_{\text{imp}}(E)$. The electron-impurity scattering is intrinsically an anisotropic mechanism but for mobility calculation it may be treated as an isotropic process by replacing the actual scattering rate by the reciprocal of the momentum relaxation time $\tau_{\text{imp}}^{-1}$ obtained by introducing the factor $1-\cos\theta$ in the integration of Eq. 8 [39]. For transport calculation in uniform material the density $n$ entering the screening length of Eq. 4 is usually chosen equal to the impurity concentration $N_{\text{imp}}$. It is not valid in the case of transport in MOS inversion layer where the local carrier density is strongly non-uniform and dependent on the applied bias. In our model the screening length is thus calculated from the local density $n(\mathbf{r})$, which is important to obtain good results for the effective mobility on the full range of effective field at any impurity concentration.

### 2.2.3. Surface-roughness scattering

In inversion layers, carrier transport can be strongly affected by irregularities at the SiO$_2$/Si interface and by charge distributions in SiO$_2$. Surface-roughness scatterings can be treated as described in Refs [18, 40] using two technology-dependent parameters $\Lambda$ and $\Delta$, the correlation length and amplitude of the interface roughness, respectively. Coulomb scattering with fixed charges located in the oxide and/ or at the interface can also be taken into account [17, 18, 41]. This approach is appropriate to a two-dimensional electron gas, but there is not any standard model for the case of three-dimensional gas. In the present Monte-Carlo simulations, only surface-roughness scattering due to the deviation of the SiO$_2$/Si interface from an ideal plane is considered and is treated with an empirical combination of specular and diffusive scattering for carriers that hit the interface [23]. Usually, the fraction of diffusive scattering $N_{\text{diff}}$ is considered as a constant value which may vary significantly, however, according to the authors: for example, $N_{\text{diff}}$ can be equal to 6% [23], 8.5% [42], 15% [24] or 50% [22]. In the present work, the fraction $N_{\text{diff}}$ has been chosen to vary with the effective vertical field $E_{\text{eff}}$ and the determination of $N_{\text{diff}}$ as a unique function of $E_{\text{eff}}$ is developed in Sec. 3.1.

### 2.2.4 . Alloy scattering in Si$_{1-x}$Ge$_x$

In the simulated structures with thin strained-Si channel some electrons may enter the Si$_{1-x}$Ge$_x$ virtual substrate where alloy scattering is to be considered. Alloy scattering is treated within the classical model of the "square well" perturbation potential of height $U_{\text{all}}$ in a sphere of volume $V$ centred on each alloy site. The radius of this sphere is arbitrarily chosen as the nearest-neighbour distance $r_0 = \sqrt{3}\, a_0/4$ where $a_0$ is the lattice parameter [43]. The alloy potential $U_{\text{all}}$ is considered equal to 0.8 eV for electrons in SiGe [44]. The isotropic alloy scattering relaxation time is given by

$$\frac{1}{\tau_{\text{all}}(E)} = \frac{3\pi\sqrt{2}\, a_0^3}{64\hbar^4} m_D^{3/2} x(1-x) U_{\text{all}}^2 (1+2\alpha E)\sqrt{E(1+\alpha E)} \qquad (9)$$

## 2.3. Simulated structures and simulation procedure

The epilayer stack of the simulated MOS structures consists of tensile-strained Si pseudomorphically grown on thick $Si_{1-x}Ge_x$ (001) pseudo-substrate. The thickness of the top strained Si layer is either 2 nm, 5 nm, 8 nm or 420 nm. The latter thickness is unrealistic for practical applications and is just considered for the purposes of comparison. The Ge content $x$ in the pseudo-substrate varies from 0 to 30%. The doping levels of Si and $Si_{1-x}Ge_x$ are identical and vary from $1\times10^{16}$ to $1\times10^{18}$ cm$^{-3}$.

Prior to Monte Carlo simulation of transport, the 1D Poisson's equation is solved in the MOS capacitor according to the gate bias $V_{GS}$ to obtain the vertical field and carrier density profiles. The corresponding effective field is deduced from Eq. 1. The Monte Carlo calculation is then performed using this vertical profile of electric field and considering a uniform parallel driving field. As mentioned in the introduction, quantization effects are not taken into account. All simulations are performed at room-temperature.

## 3. Effective mobility in bulk Si MOS structures

### 3.1. Determination of $N_{\text{diff}}$

The fraction of diffusive scattering has been determined as a function of $E_{\text{eff}}$ from an adjustment to experimental mobilities [16] obtained for an inversion Si layer in a lightly doped n-MOSFET. Indeed, for small doping levels the effects of impurity Coulomb scattering on carrier transport are not strong and the $E_{\text{eff}}$ dependence of $\mu_{\text{eff}}$ is mainly due to roughness scattering. For a given $V_{GS}$ (i.e. for a given $E_{\text{eff}}$), some simulations have been performed by varying $N_{\text{diff}}$. The obtained effective mobilities $\mu_{\text{eff}}(N_{\text{diff}})$ are then compared to the experimental $\mu_{\text{eff}}$, which allows us to deduce the relevant $N_{\text{diff}}$ value. Repeating this procedure for different $E_{\text{eff}}$ leads finally to the following expression of $N_{\text{diff}}$ as a function of $E_{\text{eff}}$ (with $E_{\text{eff}}$ in kV/cm):

$$N_{\text{diff}} = 0.176 - 2.29\times10^{-4} E_{\text{eff}} + 3.1\times10^{-7} E_{\text{eff}}^2 - 1.69\times10^{-10} E_{\text{eff}}^3 + 2.84\times10^{-14} E_{\text{eff}}^4 \qquad (10)$$

We have then assumed that this expression can be applied for any other doping levels and whatever the strain in the Si layer, which is discussed in Sec. 4.1.

### 3.2. Universal mobility curves

On Fig. 1 we have plotted, for various impurity concentrations, the $\mu_{\text{eff}}(E_{\text{eff}})$ curves obtained by Monte Carlo simulation (lines) as well as some experimental curves (lines with symbols) [11, 15, 16, 45]. As experimental mobilities, the calculated mobilities are in good agreement with the "universal mobility curve", as defined by Takagi *et al.* [16] (dashed line).

In addition, the $\mu_{eff}(E_{eff})$ curves mirror the fall-off of the experimental mobilities in weak inversion regime, which results from reduced screening of impurities at low electron density (see Eqs. 5 and 8). The respective influence of impurity and surface-roughness scatterings is clearly illustrated in Fig. 2 **where we plot the average rate of scattering for different scattering mechanisms. These rates are obtained by counting the scattering events experienced by all simulated particles (50000) for a given period of time (270 ps)**. Under high $E_{eff}$, electrons are distributed close to the SiO$_2$/Si interface, and the probability for an electron to undergo a diffusive scattering event increases for a given period of time. Indeed, as it can be seen on Fig. 2, for high $E_{eff}$ values, surface-roughness scatterings (circles) are dominant and independent of the doping level, leading to the decrease of the mobility and to the "universal curve". At low $V_{GS}$, the deviation from the "universal mobility curve" is larger as the Si doping level increases and is directly related to impurity Coulomb scattering. Indeed, Fig. 2 shows that these impurity scatterings (triangles) are obviously more frequent when the doping level is high (see the arrows on Fig. 1 and 2, indicating the points corresponding to $V_{GS}$ = 0 V). For 10$^{18}$ cm$^{-3}$, this type of scattering dominates under low $V_{GS}$ bias, while its influence decreases when increasing $V_{GS}$, as a consequence of enhanced electron density and higher screening effects. Nevertheless, for a given doping level, we can notice that the experimental curves and those obtained by simulation are not systematically superimposed at low $V_{GS}$ bias. Two possible reasons can be put forward: there is either a difference between real and announced experimental doping levels, or/and a poor estimation of screening effects for the computation of impurity scattering rate. The first explanation can be illustrated with the help of the experimental curves of Fig. 1 corresponding to doping levels of 10$^{16}$, 10$^{17}$ and 1.4×10$^{17}$ cm$^{-3}$ (triangles and reverse triangles). On one hand, for similar announced doping levels (full triangles and reverse triangles), a significant discrepancy of the $\mu_{eff}$ values is observed, and, on the other hand, identical $\mu_{eff}$ values are obtained for two different doping levels (full and open triangles). Second, each experimental curve may be fitted by calculation after adjustment of the screening length entering Eq. 8. A single approach of screening does not allow us to describe exactly each different technology, i.e. all experimental spread data found in the literature [46]. Nevertheless, the experimental trends are correctly reproduced and in addition to previous validation of phonon scattering model [4, 31], these first results validate our combined approach of Coulomb and surface-roughness scatterings. It should be mentioned again that this approach has been developed in such a way that it is easily applicable to self-consistent device Monte Carlo simulation.

## 4. Effective mobility in strained Si inversion layers

### 4.1 Effect of strain on electron mobility

The simulations have been extended to MOS structures with 8 nm-thick strained Si as surface channel on relaxed Si$_{1-x}$Ge$_x$ virtual substrate for various values of x (0.05, 0.10, 0.15, 0.20 and 0.30) and of doping level ($10^{16}$, $1.4 \times 10^{17}$, $3 \times 10^{17}$ and $10^{18}$ cm$^{-3}$). On Fig. 3, we have reported the effective mobility as a function of $E_{\text{eff}}$ for x = 0.15 and bulk Si. The curves $\mu_{\text{eff}}(E_{\text{eff}})$ obtained for the other *x* values are omitted here for clarity. Whatever *x*, these curves mirror the behaviour of $\mu_{\text{eff}}$ *versus* $E_{\text{eff}}$ obtained for bulk Si. Indeed, we observe a universal character of $\mu_{\text{eff}}$ for high effective field, as well as a drop of the mobility in weak inversion regime. We can also see the expected mobility enhancement when compared to the case of bulk Si. This effect is more clearly evidenced on Fig. 4, where we have plotted $\mu_{\text{eff}}(E_{\text{eff}})$ by varying *x* from 0 to 0.30, the doping level being equal to $10^{16}$ cm$^{-3}$. We observe an increase of the mobility with the Ge content in the pseudosubstrate, i.e. with strain in the Si inversion layer. For $x \leq 0.10$ the mobility enhancement is nearly constant over the full range of effective field. For $x \geq 0.15$, this mobility enhancement depends on $E_{\text{eff}}$: the maximum of $\mu_{\text{eff}}$ is reached for $x = 0.30$ under low $E_{\text{eff}}$ ($\leq 300$ kV/cm) and as soon as $x = 0.15$ under higher effective fields (> 300 kV/cm).

The mobility enhancement is also observed for the other doping levels studied in this work and can be explained in terms of effective mass and of **average rate of the different scattering mechanisms**. Indeed, when increasing *x*, more and more electrons reside in the $\Delta_2$ valleys of the strained Si inversion layer, and then exhibit a transverse effective mass (i.e. a low mass) in the direction of transport and a longitudinal effective mass (i.e. a high mass) in the vertical direction. Concurrently, **we observe that the average rate** of surface-roughness and intervalley phonon scatterings diminishes (see Fig. 5), which also contributes to the improvement of the mobility. The number of acoustic phonon scatterings **per unit of time** (not shown in Fig. 5) remains independent of $E_{\text{eff}}$ and strain, and is equal to about $1.93 \times 10^{12}$ s$^{-1}$. The **rate** of intervalley phonon scatterings diminishes continuously with increasing *x*, mainly due to the increase of the splitting energy $\Delta E_s$ between the $\Delta_2$ and $\Delta_4$ valleys. The decrease of **the average rate of** surface-roughness scattering with *x* is related to the increase of electron population in the $\Delta_2$ valleys. Indeed, the electrons of the $\Delta_2$ valleys that hit the SiO$_2$/Si interface have a low vertical component of velocity due to the longitudinal effective mass, and then this scattering mechanism happens less frequently when *x* increases. As soon as *x* reaches 0.15, about 96% of electrons in the strained Si are in the $\Delta_2$ valleys, and therefore increasing *x* beyond 0.15 does not almost change the number of scatterings, as it can be seen on Fig. 5 (full symbols). Under low $E_{\text{eff}}$, both intervalleys phonon and surface-roughness scattering decrease contribute to the mobility enhancement with *x*. Nevertheless, under high effective field, surface-roughness scatterings clearly dominate and the maximum of mobility is reached for $x = 0.15$, due to the saturation of the number of this scattering.

We have then compared the $\mu_{\text{eff}}(E_{\text{eff}})$ curves of Fig. 4 to experimental mobilities [7, 11-13, 45, 47]. The experimental electron mobilities *versus* $E_{\text{eff}}$ are generally extracted from the linear regime of drain current $I_D(V_{DS})$ for long surface-channel strained Si n-MOSFETs. Experimentally, the mobility enhancement is nearly constant over the full range of effective field, whatever $x$ [11, 47]: see for example the results of Currie *et al.* in the insert of Fig. 4 [11]. Rim *et al.* have shown, with the help of a simple curve-fitting analysis, that both phonon scattering rate and surface-roughness scattering rate should be reduced to correctly describe the experimental mobility in strained Si on $Si_{1-x}Ge_x$ for $x \geq 0.15$ [47]. This observation suggests that strain may influence surface-roughness scattering: indeed, Boriçi *et al.* [48] have evidenced that a $SiO_2$/ strained Si interface is less rough than a $SiO_2$/ Si one. From the point of view of our simulations, it means that considering $N_{\text{diff}}(E_{\text{eff}})$ as unchanged by strain leads to under-estimated mobility at high effective fields. Further experimental data would be necessary to check this assumption and to quantify the effect of strain on $N_{\text{diff}}$. Considering mobility values under low effective fields, typically 0.3 MV/cm, we have reported in Fig. 6 the electron mobility enhancement relative to bulk Si as a function of $x$ for our results (open symbols) and for experimental data (full symbols). The saturation of the mobility when $x$ reaches 0.15 is also evidenced experimentally (see data from Refs [11, 47]), and increasing the Ge content above 0.20 has no more effect on mobility [11]. Nevertheless, the plot of Fig. 6 shows a discrepancy between the various experimental results. For instance, for $x = 0.30$, the electron mobility can be enhanced by $\approx 80\%$ [11] or by 120% [12, 13]. These differences can be explained in terms of surface roughness of the strained $Si/Si_{1-x}Ge_x$. Sugii *et al.* investigated the relationship between Si surface morphology (periods and edge shapes of undulations) and electron mobility [49]. They obtained the best enhancement of mobility when the strained Si was grown on a pseudosubstrate consisting on a $Si_{1-y}Ge_y$ ($0 < y < x$) graded buffer layer (GBL) followed by a $Si_{1-x}Ge_x$ buffer layer, and they concluded that surface roughness must be controlled to increase mobility in strained Si layer. Recently, an approach with Monte Carlo simulations has confirmed the strong dependence of SiGe pseudosubstrate roughness (correlation length and amplitude) on mobility [34]. Other investigations showed that adding chemical-mechanical-polishing (CMP) of the pseudosubstrate improves the Si surface, whose roughness becomes of about 0.4 nm compared to about 10 nm for strained Si layer grown on GBL without CMP [12, 34]. When a thick metastable strained Si layer ($\approx 25$ nm) is then grown directly on the pseudosubstrate with CMP, mobilities are significantly enhanced when compared to devices without CMP (see full triangle [12] on Fig. 6). On the other hand, this enhancement is not observed for thin strained Si layer (8 nm in Ref. [11]) or when a regrowth of SiGe is carried out after CMP [11]. In the former case, carriers must be close to contaminants resulting from the CMP, and in the latter case the roughness has again increased after regrowth of SiGe. According to Fig. 6, a large enhancement of $\mu_{\text{eff}}$ is also observed for Refs [13] and [7] (full triangle and square, respectively), despite no CMP has been used. Olsen *et al.* obtained low surface roughness by growing a $Si/Si_{0.7}Ge_{0.3}$ structure on a relaxed constant composition $Si_{0.85}Ge_{0.15}$ virtual substrate [13] and the same mobility enhancement as Sugii *et al.* was achieved [12]. Concerning the result of Rim *et al.*, the lack of experimental details does not allow us to explain this mobility enhancement [7]. For the other data plotted on this figure, the technological processes for NMOS fabrication including a GBL were similar and lead to close mobility values [11, 45, 47]. Our results obtained from Monte Carlo simulation are in accordance with the best experimental mobility enhancements [7, 12, 13].

**Other modelling approaches have considered the quantized nature of the inversion layer for transport calculation in strained Si [40, 41, 50, 51]. Using the standard Ando's model of roughness scattering [40] it has been shown that the scattering rate in strained Si is strongly overestimated [41]. The model has been improved by Gámiz et al. [50,51] to include the effect of the buried layer (SiGe or $SiO_2$) on the perturbation Hamiltonian, which yields satisfactory mobility behaviour versus transverse electric field [51]. It should be noted that if plotted on a linear scale, the results obtained from our simple empirical approach are in good agreement with that of Ref. 51. A similar method for the case of a 3D electron gas has been developed by including the correlation length $\Lambda$ and the amplitude $\Delta$ of surface fluctuations in the determination of the specular scattering probability which, however, is independent of transverse field [52]. Using this approach, a correct agreement with experimental high-field mobility in strained Si is found only if $\Lambda$ and $\Delta$ are significantly longer and smaller, respectively, than in relaxed Si, which is an additional indication that the strain in Si reduces the roughness of the $SiO_2$ interface.**

## 4.2 Effect of strained Si thickness on electron mobility

We have also investigated the effect of the strained Si thickness on electron mobility. Keeping the Ge content and the doping level constant (0.15 and $10^{16}$ cm$^{-3}$, respectively), simulations with four different Si thicknesses have been performed. The corresponding $\mu_{\text{eff}}(E_{\text{eff}})$ curves are plotted on Fig. 7. An influence of the strained Si thickness is obvious under low effective field. Indeed, $\mu_{\text{eff}}$ falls off when decreasing the Si thickness down to 5 nm. Between 5 nm and 420 nm, the values of mobility are quite close to each other. Under high effective fields, the mobility remains unchanged whatever the strained Si thickness. In these structures, the distribution of electrons in the different layers as a function of $E_{\text{eff}}$ (see full symbols of Fig. 8) show that under low effective field, the percentage of electrons present in the SiGe pseudosubstrate increases with decreasing the strained Si thickness. This leads to parasitic electron conduction through the low-mobility SiGe underlayer, which degrades the overall mobility. Additionally, the number of electrons close to the $SiO_2$/Si interface is higher for thinner strained Si layers, which is then accompanied by an increase of the **average rate** of surface-roughness scattering events (Fig. 8, open symbols). These facts explain why, under low effective field, mobility decreases with decreasing the Si thickness. Under high effective fields, the electrons are all located in the strained Si layer and the **average rate of surface-roughness scattering** is identical from a structure to another. Whatever the strained Si thickness, 96% of the electrons located in the Si exhibit a transverse mass, and therefore the Si thickness has no influence on the mobility, as it can be seen on Fig. 7. Currie *et al*. have studied the effect of channel thickness on mobility enhancement for NMOS devices [11]. They observed large mobility degradations for the thinnest channels, and as soon as the channel thickness increases beyond 5 nm, the devices exhibit the expected mobility enhancements we have reported on Fig. 6 (full circles). This trend is in accordance with our observations.

**Conclusion**

In this work, we have carefully analysed the effect of strain in Si, doping level and Si thickness on electron transport in the inversion layer of a strained Si/ $Si_{1-x}Ge_x$ MOS structure. Our investigations were based on the calculation of effective mobility versus effective field using Monte Carlo simulations. As for bulk Si inversion layers, a universal relationship between $\mu_{eff}$ and $E_{eff}$ is found when Si is strained. In addition, we observe the deviation from these universal curves in weak inversion regime, due to Coulomb scatterings with impurities. The experimental trends are correctly reproduced by these results, which validates our approach of phonon, Coulomb and surface-roughness scatterings. Regarding the strain-induced mobility enhancement factor (maximum of 120 % for strained Si/ $Si_{0.70}Ge_{0.30}$), our simulation results match well with best available experimental data. The role of Si thickness is evidenced under low effective field and for thickness lower than 5 nm. A degradation of the mobility is then observed, due to electron conduction through the SiGe underlayer and increase of surface-roughness scatterings.

**Acknowledgments**

This work was partially supported by the European Community 6th FP (Network of Excellence SINANO, contract no 506844).

# References

[1] Van de Walle CG. Strain effect on the valence-band structure of SiGe. In: Kasper E, editor. Properties of strained and relaxed silicon germanium, London: INSPEC, EMIS Datareviews Series 12; 1995. p. 94-102.

[2] Ismail K, Nelson SF, Chu JO, Meyerson BS. Electron transport properties of Si/SiGe heterostructures: measurements and device implications. Appl. Phys. Lett. 1993;63:660-2.

[3] Hackbarth T, Hoeck G, Herzog HJ, and Zeuner M. Strain relieved SiGe buffers for Si-based heterostructure field-effect transistors. J. Cryst. Growth 1999;201/202:734-38.

[4] Monsef F, Dollfus P, Galdin-Retailleau S, Herzog HJ, Hackbarth T. Electron transport in Si/SiGe modulation-doped heterostructures using Monte Carlo simulation. J. Appl. Phys. 2004;95:3587-93.

[5] Zeuner M, Hackbarth T, Enciso-Aguilar M, Aniel F, von Känel H. Sub-100 nm gate technologies for Si/SiGe buried channel RF devices. Jpn. J. Appl. Phys. 2003;42:2363-5.

[6] Aniel F, Enciso-Aguilar M, Giguerre L, Crozat P, Adde R, Mack T, Seiler U, Hackbarth T, Herzog HJ, König U, Raynor B. High performance 100 nm T-gate strained $Si/Si_{0.6}Ge_{0.4}$ n-MODFET. Solid-State Electron. 2003;47:283-9.

[7] Rim K, Chu J, Chen H, Jenkins KA, Kanarsky T, Lee K, Mocuta A, Zhu H, Roy R, Newbury J, Ott J, Petrarca K, Mooney P, Lacey D, Koester S, Chan K, Boyd D, Ieong M, Wong HS. Characteristics and device design of sub-100 nm strained Si N- and PMOSFETs. In: 2002 Symposium on VLSI technology Digest, New York: IEEE; 2002. p. 98-9.

[8] Olsen SH, O'Neill AG, Chattopadhyay S, Kwa KSK, Driscoll LS, Norris DJ, Cullis AG, Robbins DJ, Zhang J. Evaluation of strained Si/ SiGe material for high performance CMOS. Semicond. Sci. Technol. 2004;19:707-14.

[9] Kim K, Chuang C-T, Rim K, Joshi RV. Performance assessment of scaled strained-Si channel-on-insulator (SSOI) CMOS. Solid-State Electron. 2004;48:239-43.

[10] Langdo TA, Currie MT, Cheng Z-Y, Fiorenza JG, Erdtmann M, Braithwaite G, Leitz CW, Vineis CJ, Carlin JA, Lochtefeld A, Bulsara MT, Lauer I, Antoniadis DA, Somerville M. Strained Si on insulator technology: from materials to devices. Solid-State Elect. 2004;48:1357-67.

[11] Currie MT,. Leitz CW, Langdo TA, Taraschi G, Fitzgerald EA, Antoniadis DA. Carrier mobilities and process stability of strained Si n- and p-MPOSFETs on SiGe virtual substrate. J. Vac. Sci. Technol. B 2001;19:2268-79.

[12] Sugii N, Hisamoto D, Washio K, Yokoyama N, Kimura S. Performance enhancement of strained-Si MOSFETs fabricated on a chemical-mechanical-polished SiGe substrate. IEEE Trans. Electron Dev. 2002;49:2237-43.

[13] Olsen SH, O'Neill AG, Driscoll LS, Kwa KSK, Chattopadhyay S, Waite AM, Tang YT, Evans AGR, Norris DJ, Cullis AG, Paul DJ, Robbins DJ. High-performance nMOSFETs

using a novel strained Si/SiGe CMOS architecture. IEEE Trans. Electron Dev. 2002;50:1961-8.

[14] Sabnis AG, Clemens JT. Characterization of the electron mobility in the inverted <100> Si surface. In: IEDM Tech. Dig. 1979, p.18-21

[15] Sun SC, Plummer JD. Electron mobility in inversion and accumulation layers on thermally oxidized silicon surfaces. IEEE Trans. Electron Dev. 1980;27:1497-508.

[16] Takagi S, Toriumi A, Iwase M, Tango H. On the universality of inversion layer mobility in Si MOSFET's: Part I-Effects of substrate impurity concentration. IEEE Trans. Electron Dev. 1994;41:2357-62.

[17] Gámiz F, Melchor I, Palma A, Cartujo P, López-Villanueva JA. Effects of oxide-charge space correlation on electron mobility in inversion layers. Semicond. Sci. Technol. 1994;9:1102-7.

[18] Fischetti MV, Laux SE. Monte Carlo study of electron transport in silicon inversion layers. Phys. Rev. B 1993;48:2244-74.

[19] Villa S, Lacaita AL. A physically-based model of the effective mobility in heavily-doped n-MOSFET's. IEEE Trans. Electron Dev. 1998;45:110-5.

[20] Esseni D., Sangiorgi E. Low field electron mobility in ultra-thin SOI MOSFETs: experimental characterization and theoritical investigation. Solid-State Elect. 2004;48:927-36.

[21] Dollfus P, Bournel A, Galdin S, Barraud S, Hesto P. Effect of discrete impurities on electron transport in ultra-short MOSFET using 3D Monte Carlo simulation. IEEE Trans. Electron Dev. 2004;51:749-56.

[22] Fischetti MV, Laux SE. Monte Carlo simulation of transport in technologycally significant semiconductors of the diamond and zinc-blende structures- Part II: submicrometer MOSFET's. IEEE Trans. Electron Dev. 1991;38:650-60.

[23] Sangiorgi E, Pinto MR. A semi-empirical model of surface scattering for Monte Carlo simulation of silicon n-MOSFET's. IEEE Trans. Electron Dev. 1992;39:356-61.

[24] Bufler FM, Fichtner W. Scaling of strained-Si n-MOSFETs into the ballistic regime and associated anisotropic effects. IEEE Trans. Electron Dev. 2003;50:278-84.

[25] Richard S, Cavassilas N, Aniel F, Fishman G. Strained silicon on SiGe: temperature dependence of carrier effective masses. J. Appl. Phys. 2003;94:5088-94.

[26] Fischetti MV, Laux SE. Band structure, deformation potentials, and carrier mobility in strained Si, Ge and SiGe alloys. J. Appl. Phys. 1996;80:2234-52.

[27] Galdin S, Dollfus P, Aubry-Fortuna V, Hesto P, Osten HJ. Band offsets predictions for strained group IV alloys: $Si_{1-x-y}Ge_xC_y$ on Si (001) and $Si_{1-x}Ge_x$ on $Si_{1-z}Ge_z$ (001). Semicond. Sci. Technol. 2000;15:565-72.

[28] Dollfus P, Galdin S, Osten HJ, Hesto P. Band offsets and electron transport calculation for strained $Si_{1-x-y}Ge_xC_y$/Si heterostructures. Materials Science: Materials in Electronics 2001;12:245-8.

[29] Jacoboni C, Lugli P. The Monte Carlo method for semiconductor device simulation. Wien: Springer-Verlag; 1989.

[30] Ferry DK. First order optical and intervalley scattering in semiconductors. Phys. Rev. B 1976;14:1605-9.

[31] Dollfus P. Si/Si$_{1-x}$Ge$_x$ heterostructures: electron transport and field-effect transistor operation using Monte Carlo simulation. J. Appl. Phys. 1997;82:3911-6.

[32] Yamada T, Zhou JR, Miyata H, Ferry DK. In-plane transport properties of Si/Si$_{1-x}$Ge$_x$ structure and its FET performance by computer simulation. IEEE Trans. Electron Devices1994;41:1513-22.

[33] Formicone GF, Saraniti M, Vasileska DZ, Ferry DK. Study of a 50-nm nMOSFET by ensemble Monte Carlo simulation including a new approach to surface roughness and impurity scattering in the Si inversion layer. IEEE Trans. Electron Devices 2002;49:125-32.

[34] Kitagawa I, Maruizumi T, Sugii T. Theory of electron-mobility degradation caused by roughness with long correlation length in strained-silicon devices. J. Appl. Phys. 2003;94:465-70.

[35] Takimoto N. On the screening of impurity potential by conduction electrons. J. Phys. Soc. Jpn. 1959;14:1142-58

[36] Joshi RP, Ferry DK. Effect of multi-ion screening on the electronic transport in doped semiconductors: a molecular-dynamics approach. Physical Review B 1991;43:9734-9.

[37] Hall GL. Ionized impurity scattering in semiconductors. J. Phys. Chem. Solids 1962;23:1147-51.

[38] Abramowitz M, Stegun IA. Handbook of Mathematical Functions. 9th ed. New York: Dover Publications Inc.; 1970.

[39] Kosina H. Efficient evaluation of ionized-impurity scattering in Monte Carlo transport calculations. Phys. Stat. Sol. (a) 1997;163:475-89.

[40] Ando T, Fowler AB, Stern F. Electron properties of two-dimensional systems. Rev. Mod. Phys. 1982;54:437-672.

[41] Fischetti MV, Gámiz F, Hänsch W. On the enhanced electron mobility in strained-silicon inversion layers. J. Appl. Phys. 2002;92:7320-4.

[42] Fiegna C, Sangiorgi E. Modeling of high-energy electrons in MOS devices at the microscopic level. IEEE Trans. Electron Dev. 1993;40:619-27.

[43] Harrison JW. Alloy scattering in ternary III-V compounds. Phys. Rev. B 1976;13:5347-50.

[44] V. Venkataraman, C. W. Liu, and J. C. Sturm. Alloy scattering limited transport of two-dimensional carriers in strained Si$_{1-x}$Ge$_x$ quantum wells. Appl. Phys. Lett. 1993;63: 2795-7.

[45] Rim K, Hoyt JL, Gibbons JF. Fabrication and analysis of deep submicron strained-Si N-MOSFET's. IEEE Trans. Electron Dev. 2000;47:1406-15.

[46] Roldán JB, Gámiz F. Simulation and modelling of transport properties in strained-Si and strained-Si/SiGe-on-insulator MOSFETs. Solid-State Electron. 2004;48:1347-55.

[47] Rim K, Anderson R, Boyd D, Cardone F, Chan K, Chen H, Christansen S, Chu J, Jenkins K, Kanarsky T, Koester S, Lee BH, Lee K, Mazzeo V, Mocuta A, Mocuta D, Mooney PM, Oldiges P, Ott J, Ronsheim P, Roy R, Steegen A, Yang M, Zhu H, Ieong M, Wong HSP. Strained Si CMOS (SS CMOS) technology: opportunities and challenges. Solid-State Elect. 2003;47:1133-9.

[48] Boriçi M, Watling JR, Wilkins RCW, Yang L, Barker JR. Interface roughness scattering and its impact on electron transport in relaxed and strained Si n-MOSFETs. Semicond. Sci. Technol. 2004;19:S155-7.

[49] Sugii N, Nakagawa K, Yamaguchi S, Miyao M, Role of $Si_{1-x}Ge_x$ buffer layer on mobility enhancement in a strained-Si *n*-channel metal-oxide semiconductor field-effect transistor. Appl. Phys. Lett. 1999;75:2948-50.

**[50] Gámiz F, Roldán JB, López-Villanueva JA, Cartujo-Cassinello P, Carceller JE, J. Appl. Phys. 1999, 86:6854-63.**

**[51] Gámiz F, Cartujo-Cassinello P, Roldán JB, Jiménez-Molinos F, J. Appl. Phys. 2002, 92:288-95.**

**[52] Watling JR, Yang L, Boriçi M, Wilkins RCW, Asenov A, Barker JR, Roy S, Solid-State Electron. 2004, 48:1337-46.**

**Figure captions**

Fig. 1: Calculated and experimental [11, 15, 16, 45] $\mu_{eff}(E_{eff})$ curves for electrons in bulk Si inversion layers for various doping levels (in cm$^{-3}$). Arrows indicate the bias point $V_{GS}=0$.

Fig. 2: **Average rate** of surface-roughness scattering (circles), intervalleys phonon scattering (rhombus) and Coulomb scattering with doping impurities (triangles) as a function of $E_{eff}$ in two different doped bulk Si: $1\times10^{16}$ cm$^{-3}$ (open symbols) and $1\times10^{18}$ cm$^{-3}$ (full symbols). **These rates are obtained by counting the scattering events experienced by all simulated particles (50000) for a given period of time (270 ps).**

Fig. 3: Calculated $\mu_{eff}(E_{eff})$ curves for electrons in 8 nm-thick strained Si/ Si$_{0.85}$Ge$_{0.15}$ and bulk Si inversion layers for various doping levels (cm$^{-3}$).

Fig. 4: Calculated $\mu_{eff}(E_{eff})$ curves as a function of $x$ for electrons in 8 nm-thick strained Si/Si$_{1-x}$Ge$_x$ inversion layers doped at $1\times10^{16}$ cm$^{-3}$. For comparison, experimental results of Currie *et al.* [11] are reported in the insert.

Fig. 5: **Average rate** of surface-roughness scattering (full symbols) and intervalleys phonon scattering (open symbols) as a function of $E_{eff}$ and for various $x$ values in 8 nm-thick strained Si/Si$_{1-x}$Ge$_x$ inversion layers doped at $1\times10^{16}$ cm$^{-3}$.

Fig. 6: Electron mobility enhancement deduced from our simulations (open circles) and from experimental results (full symbols) [7, 11-13, 45, 47].

Fig. 7: Calculated $\mu_{eff}(E_{eff})$ curves as a function of Si thickness for electrons in strained Si/Si$_{0.85}$Ge$_{0.15}$ inversion layers doped at $1\times10^{16}$ cm$^{-3}$.

Fig. 8: **Average rate** of surface-roughness scatterings as a function of $E_{eff}$ ans Si thickness in strained Si/Si$_{0.85}$Ge$_{0.15}$ inversion layers doped at $1\times10^{16}$ cm$^{-3}$ (open symbols). The percentage of electrons in the Si$_{0.85}$Ge$_{0.15}$ underlayer is also reported (full symbols).

# Tables

Table I. Material and phonon scattering parameter for electrons in Si

| | |
|---|---|
| Lattice constant $a_0$ (Å) | 5.431 |
| Density $\rho$ (g/cm$^3$) | 2.329 |
| Longitudinal sound velocity $u_l$ (cm/s) | $9.0 \times 10^5$ |
| Transverse sound velocity $u_t$ (cm/s) | $5.4 \times 10^5$ |
| Dielectric constant $\varepsilon_s$ | 11.7 |

*Intravalley acoustic phonon scattering:*

| | |
|---|---|
| Deformation Potential $D_{ac}$ (eV) | 6.6 |

*Intervalley phonon scattering:*

| phonon energy $\hbar\omega_{iv}$ (meV) | Deformation Potential $D_0$ (eV/cm) | $D_1$ (eV) | Scattering mode (type) |
|---|---|---|---|
| 11.4 | | 3.0 | TA (g1) |
| 18.8 | | 3.0 | LA (g2) |
| 63.2 | $3.4 \times 10^8$ | | LO (g3) |
| 21.9 | | 3.0 | TA (f1) |
| 46.3 | $3.4 \times 10^8$ | | LA (f2) |
| 59.1 | $3.4 \times 10^8$ | | TO (f3) |

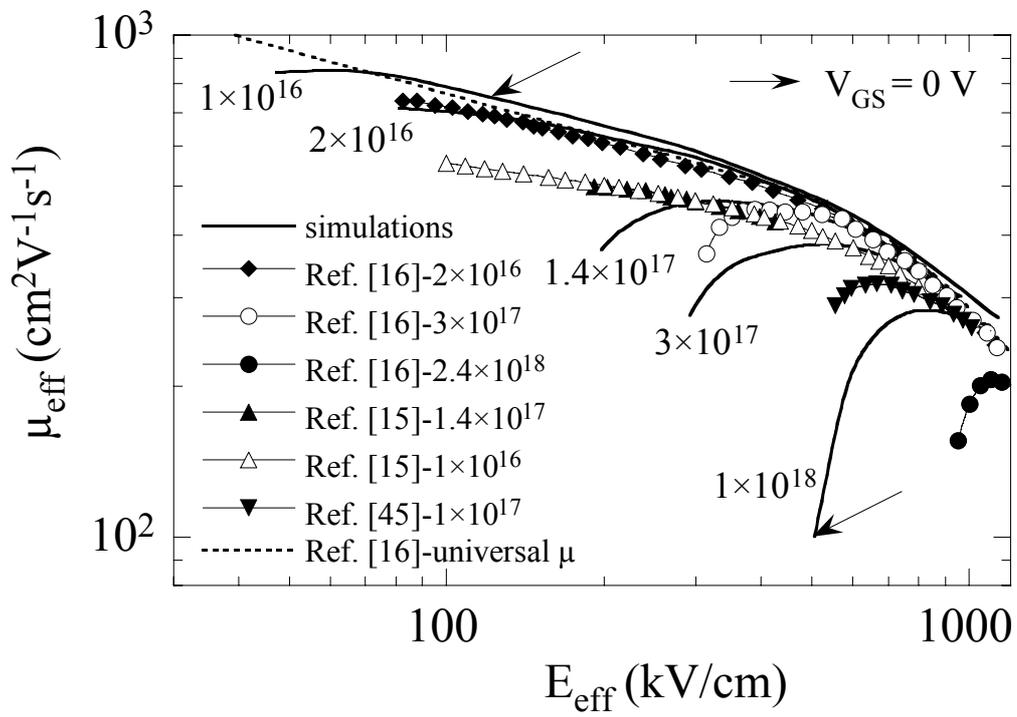

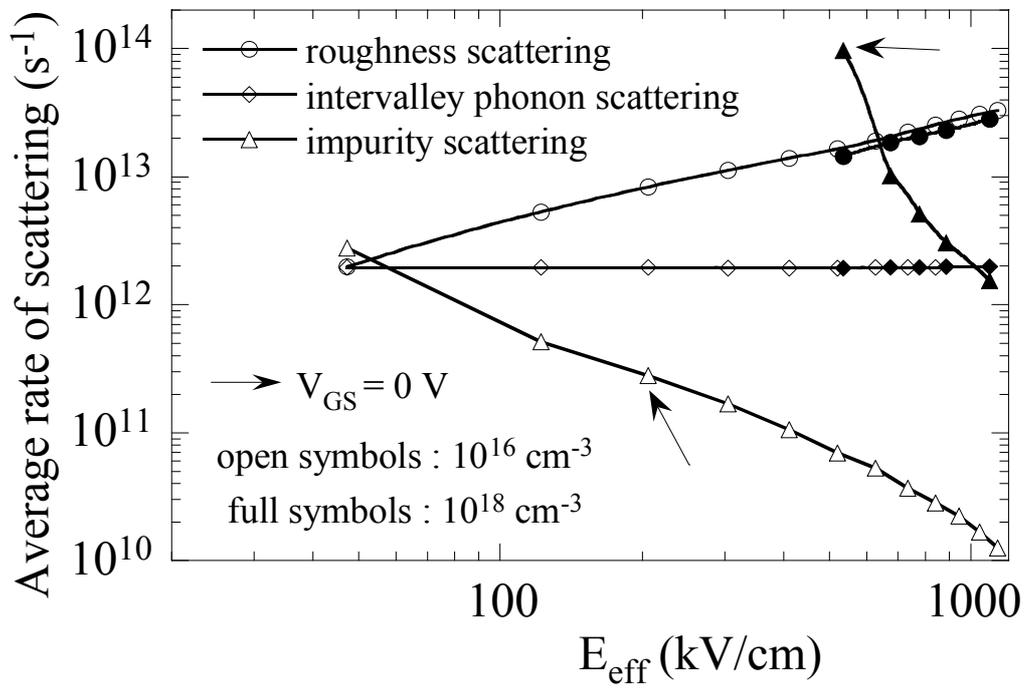

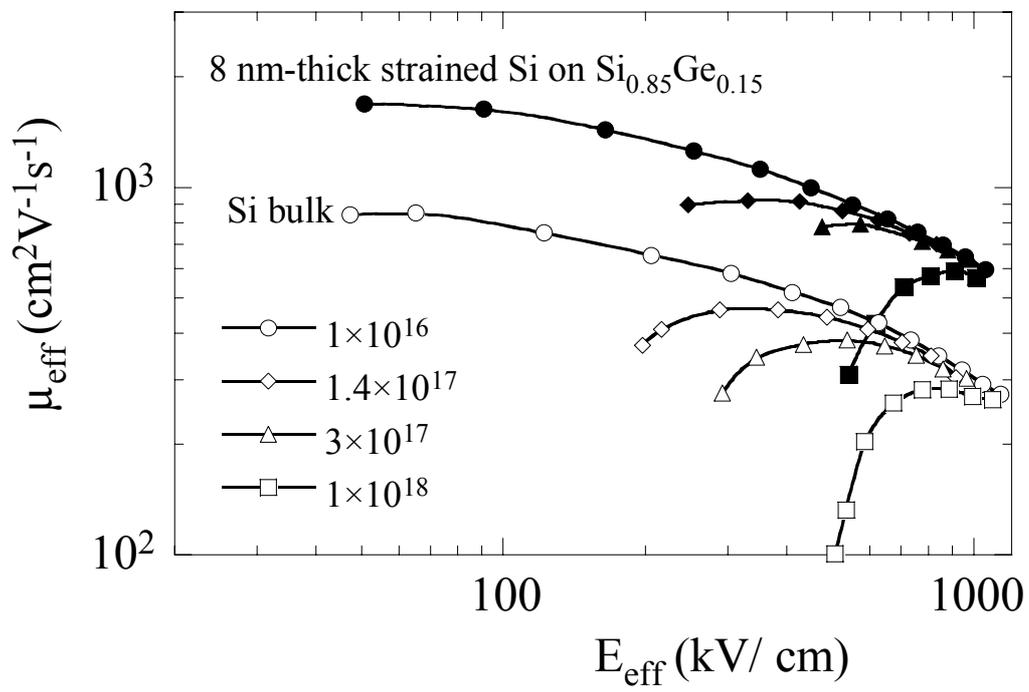

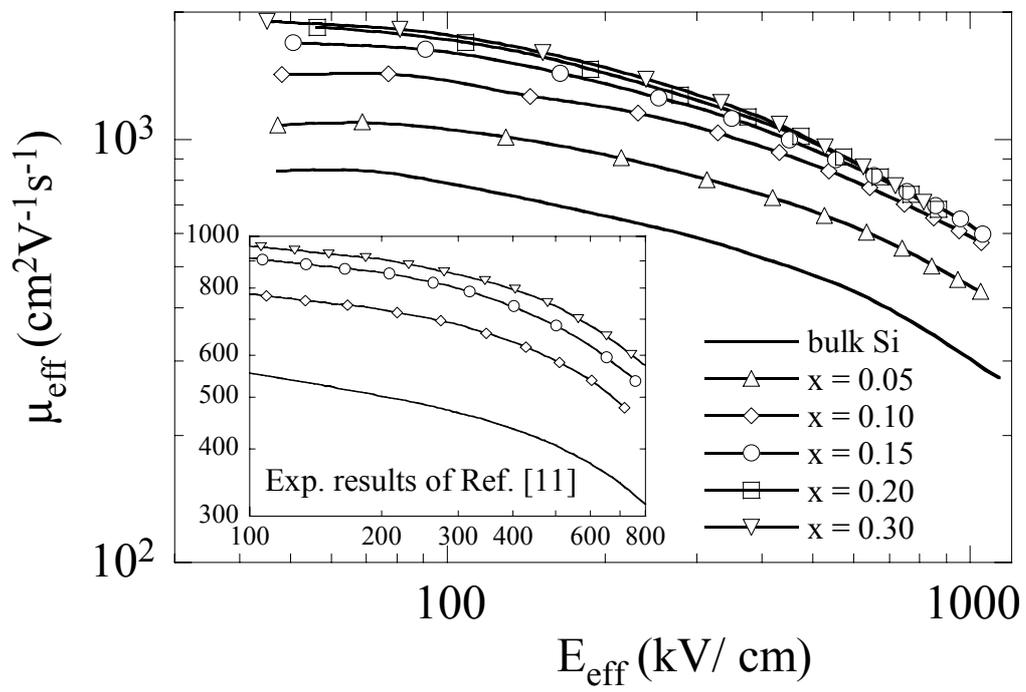

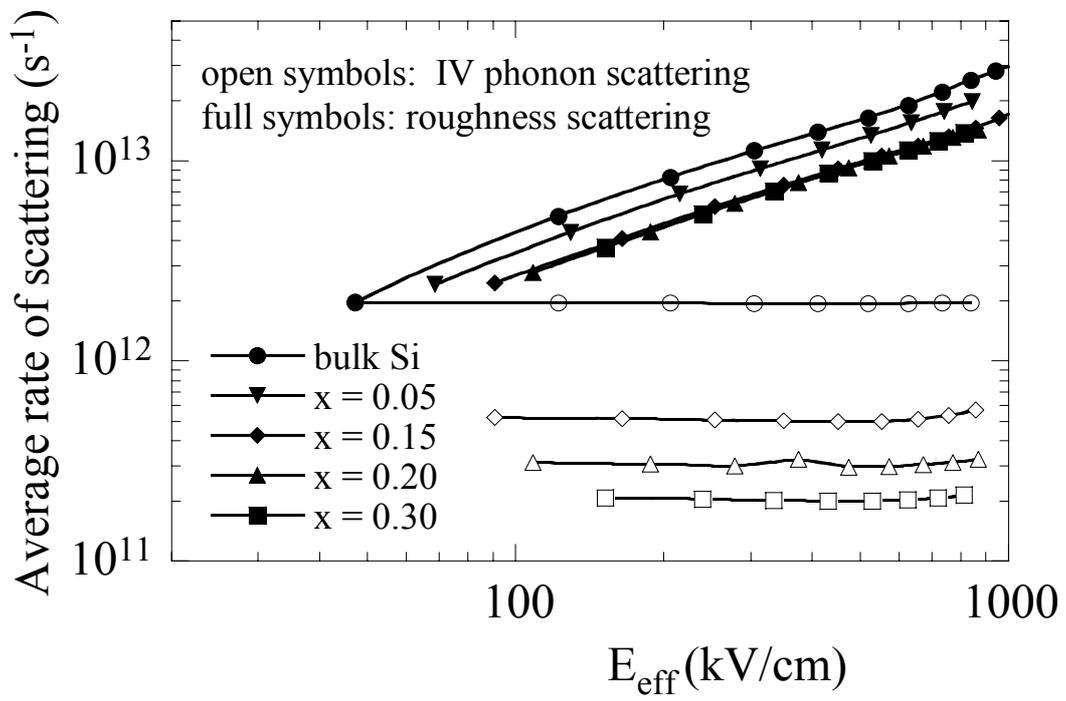

Aubry-Fortuna et al.          **Figure 6**Aubry-Fortuna et al. **Figure 6**

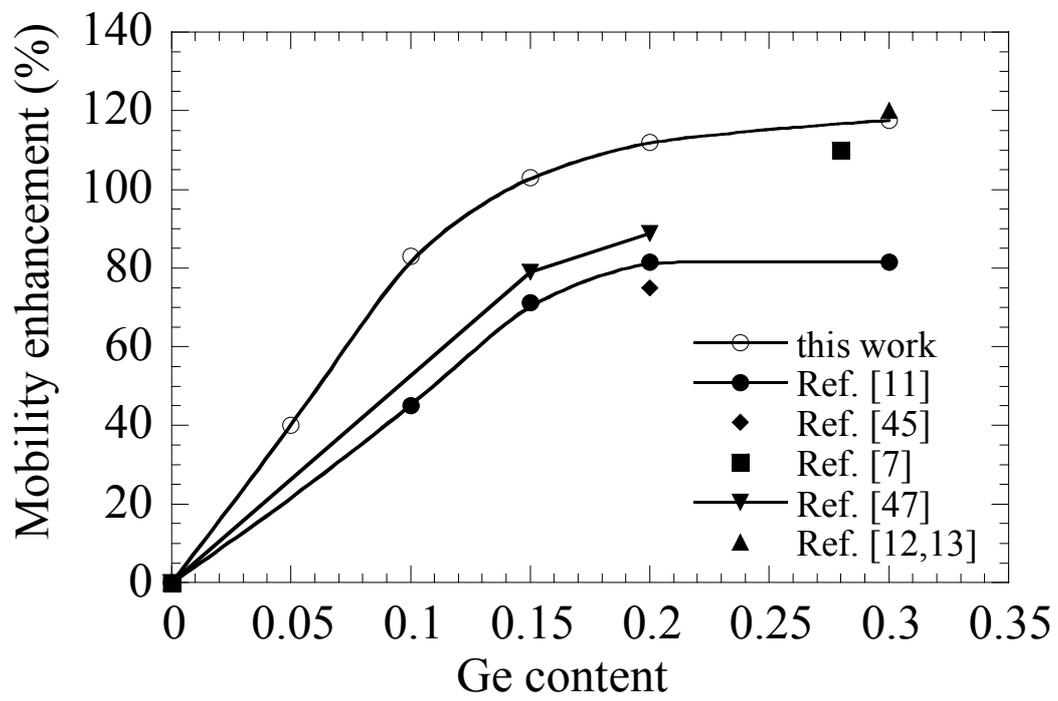

-24--24-

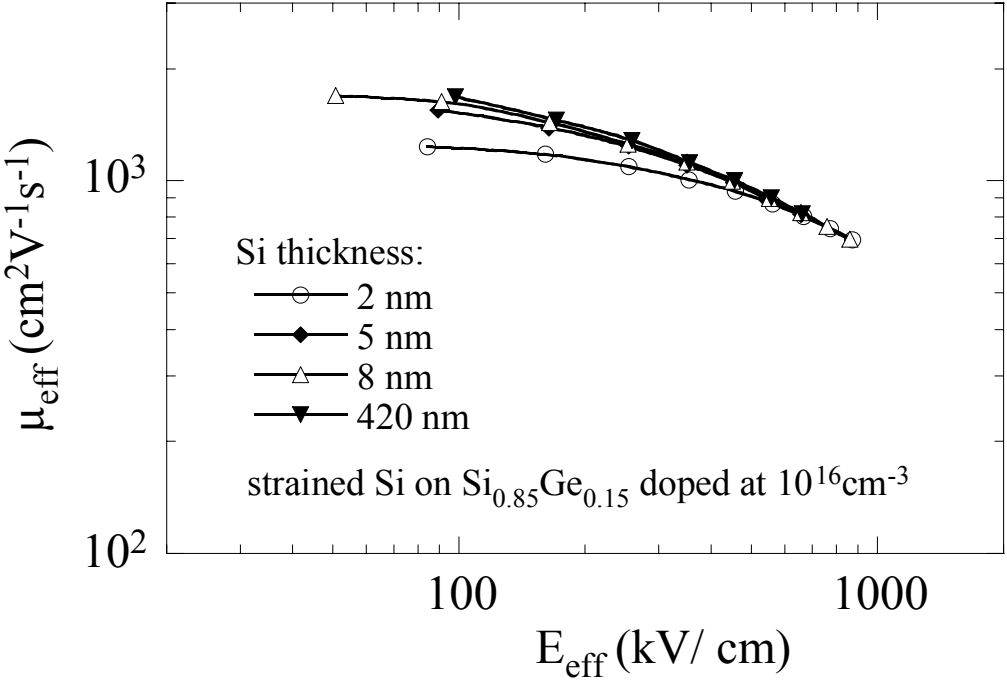

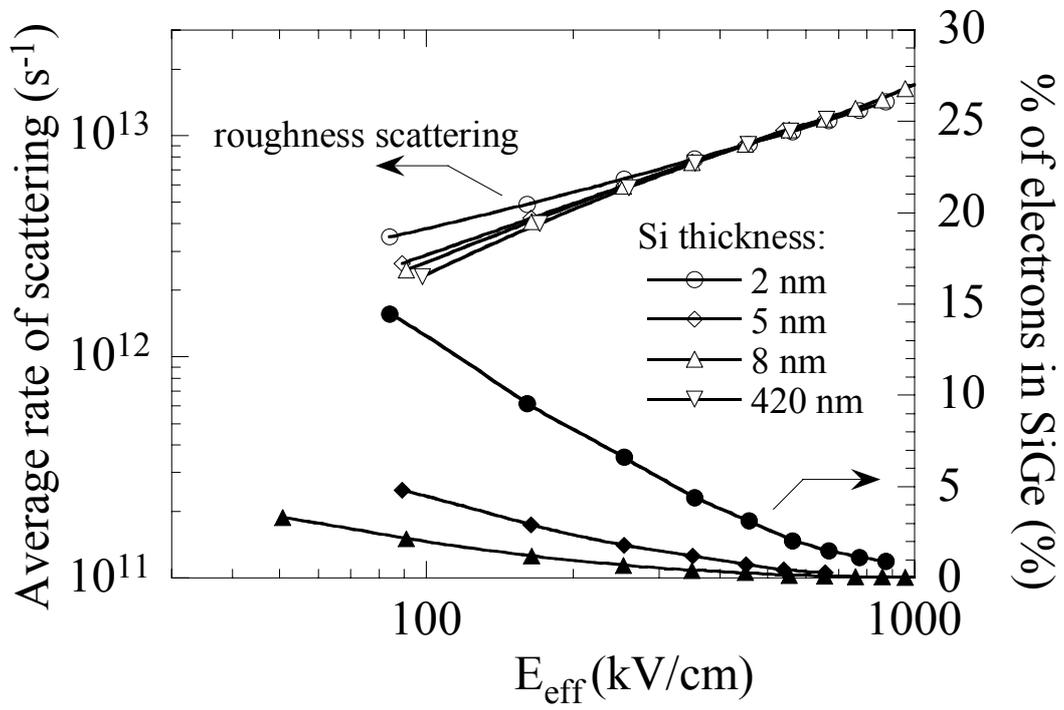